\documentclass[aps,pre,twocolumn,floatfix,superscriptaddress,showpacs]{revtex4}
\usepackage{graphicx,amsmath}
\begin{document}
\title{Asymmetric Wave Propagation in Nonlinear Systems}
\author{Stefano Lepri}
\affiliation{CNR-Consiglio Nazionale delle
Ricerche, Istituto dei Sistemi Complessi, 
via Madonna del piano 10, I-50019 Sesto Fiorentino, Italy}
\author{Giulio Casati}
\affiliation{Center for
Nonlinear and Complex Systems, Universit\`a degli Studi
dell'Insubria, Como, Italy} \affiliation{Istituto
Nazionale di Fisica Nucleare, Sezione di Milano, Milan, Italy }
\affiliation{Centre for Quantum Technologies, National University
of Singapore, Singapore 117543}
\date{\today}
\begin{abstract}
A mechanism for asymmetric (nonreciprocal) wave transmission is presented.
As a reference system, we consider a layered nonlinear, non mirror-symmetric 
model described by the one-dimensional 
Discrete Nonlinear Schr\"odinger equation with spatially varying 
coefficients embedded in an otherwise linear 
lattice. We construct a class of exact extended solutions such that
waves with the same frequency and incident amplitude impinging from 
left and right directions have very different transmission coefficients.
This effect arises already for the simplest case of two nonlinear 
layers and is associated with the shift of nonlinear resonances.
Increasing the number of layers considerably increases the complexity
of the family of solutions. Finally, numerical simulations of asymmetric 
wavepacket transmission are presented which beautifully display the 
rectifying effect.
\end{abstract}
\pacs{05.45.-a}
\maketitle


The design of small devices to control energy and/or  mass flows at different
spatial scales is a suggestive challenge from both a theoretical and applied
viewpoint. Since the first  proposal of a thermal diode \cite{Terraneo02},
capable of transmitting  heat asymmetrically between two temperature sources,
several studies appeared in the literature \cite{Segal04}, including some  experimental
realizations \cite{Chang06,Kobayashi09}.

A related issue concerns the possibility of devising a ``wave diode"  in which
electromagnetic or elastic waves are transmitted differently along two opposite
propagation directions. A so-called optical diode has been  
proposed in Ref.~\cite{Scalora94,Tocci95} and later on discussed both 
theoretically \cite{Konotop02,Liang09} and experimentally \cite{Gallo01}. 
There is also a recent proposal for a diode based on
left-handed metamaterials \cite{Feise05}. 
Another domain of application is the propagation of acoustic pulses in granular
systems. Indeed, experimental studies demonstrated a change of solitary wave
reflectivity from the interface of two granular media \cite{Nesterenko05}.

The basic idea for a ``wave diode" was proposed for
nonlinear photonic crystals \cite{Konotop02}: the rectification depends on 
whether the second harmonic of the fundamental wave is transmitted or not.
As a definition of a rectifying device we would instead propose 
that the transmitted  power at fixed incident amplitude and at the 
\textit{same frequency $\omega$}
should be sensibly different in the two opposite propagation directions.
In the present Letter we pursue a novel approach that exploits distinctive
features of nonlinear dynamical systems such as multistabiliy and
amplitude-dependent resonances to achieve such an effect. 
  
\begin{figure}[ht]
\begin{center}
\includegraphics[width=0.35\textwidth,clip]{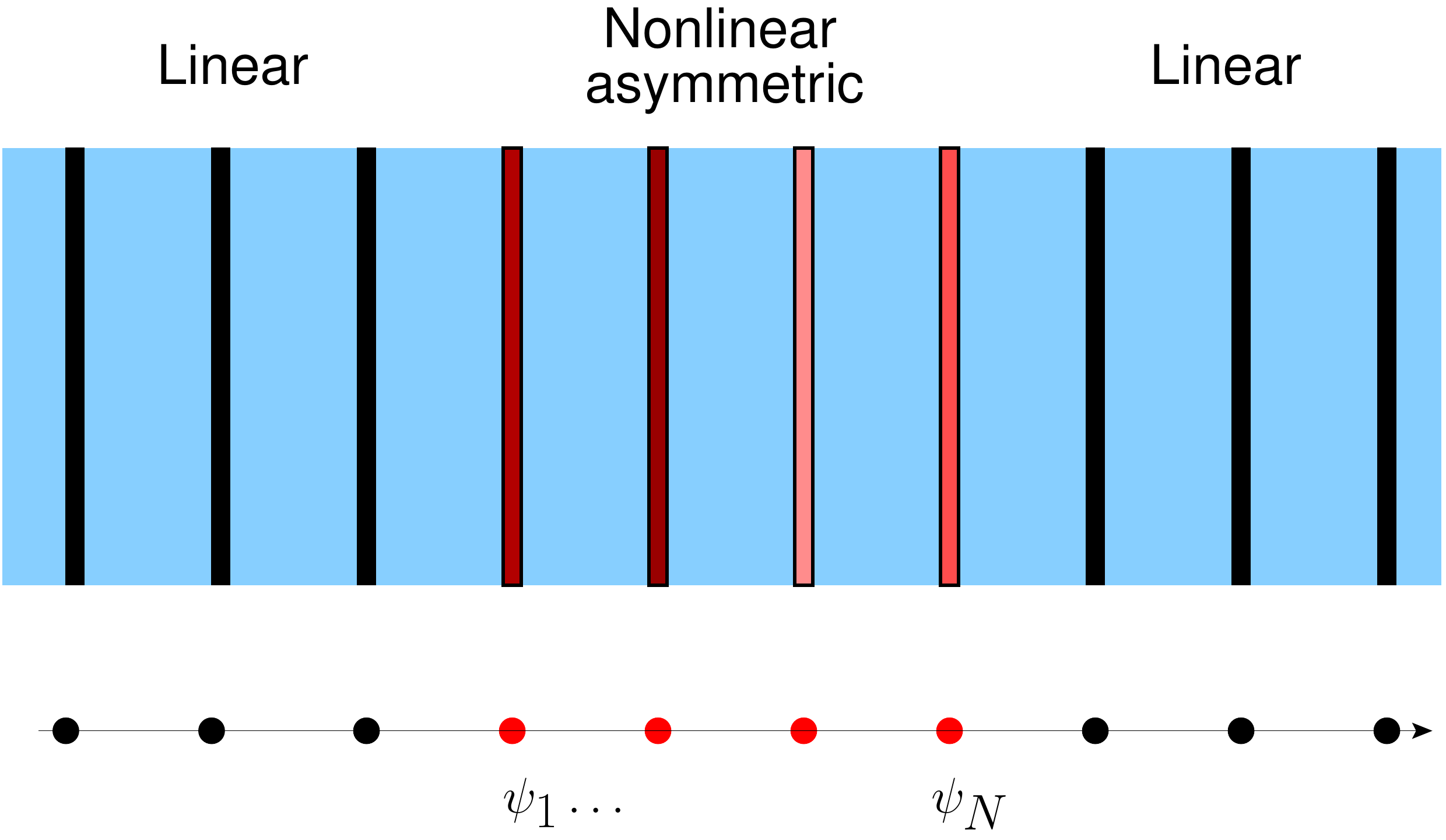}
\caption{(Color online) Sketch of a layered photonic or phononic system.
The central $N$ layers are nonlinear, non mirror-symmetric with respect
to the structure center.
In the limit of a vanishing width of the thinner layers the dynamics
within high-frequency Bloch bands is described by a DNLS equation 
for the envelope $\psi_n$ (see Refs.  \cite{Kosevich02,Hennig99} for
details).
}
\label{f:fig0}
\end{center}
\end{figure}

In a linear, time-reversal symmetric system this possibility 
is forbidden by the reciprocity theorem 
\cite{Rayleigh}. Therefore, 
one needs to consider \textit{nonlinear and asymmetric} systems 
\cite{Narayan04}. As a reference model we
will focus on the Discrete Nonlinear Schr\"odinger (DNLS) equation 
\cite{Eilbeck1985,Kevrekidis} with spatially varying coefficients. It has been
demonstrated \cite{Kosevich02} that DNLS can be a sensible  approximation for
the evolution of longitudinal Bloch waves in layered photonic or phononic
crystals (Fig.~\ref{f:fig0}). Variable  coefficients describe different nonlinear properties of each
layer and the presence of defects. Beyond its relevance in many different
physical contexts, the DNLS equation has the  big advantage of being among the
simplest dynamical systems  amenable to a complete theoretical analysis. For our
purpose,  it is particularly convenient as it allows to solve the scattering 
problem exactly without the complications of having to deal with  wave
harmonics.

More precisely, let us consider the stationary DNLS 
equation defined on an infinite one-dimensional lattice
\begin{equation}
\omega \psi_n = V_n\psi_n 
-\psi_{n+1} - \psi_{n-1} + \alpha_n |\psi_n|^2 \psi_n \qquad.
\label{dnls}
\end{equation}
We will assume the usual scattering setup where $V_n$ and $\alpha_n$ 
are nonvanishing only for $1\le n \le N $. The two semiinfinite portions 
($n<1$, $n>N$) of the lattice, model two leads where the wave can 
propagate freely. 

Let us look for solutions of the associated transmission problem
\begin{equation}
\psi_n = 
\begin{cases}
R_0 e^{ikn} + R e^{-ikn} & n\le 1 \\
T e^{ikn}                & n \ge N
\end{cases}
\label{wave}
\end{equation}
where $\omega = -2 \cos k$ and $0\le k \le \pi$ for the wave coming
from the left direction; $R_0, R$ and $T$
are the incident, reflected and transmitted amplitudes respectively.
The  solution sought must be complex in order to carry 
a nonvanishing current $J=2|T|^2\sin k$. 

To break the mirror symmetry with respect to the center of the 
nonlinear portion, one must choose at least 
one of the two sets of coefficients $V_n$, $\alpha_n$
such that $V_n\ne V_{N-n+1}$, $\alpha_n\ne\alpha_{N-n+1}$. 
Note that the transmission of the right-incoming wave with the 
same $R_0$ and $\omega$ is computed by 
solving the problem with $(V_n,\alpha_n)\longrightarrow 
(V_{N-n+1},\alpha_{N-n+1})$ (i.e. ``flipping  the sample").
In the following, we will refer to such solutions as those
having negative wavenumbers, $-k$. 
Nonlinearity is essential as for $\alpha_n=0$ the transmission 
coefficient is the same for waves 
coming from the left or right side, independently on $V_n$
due to time-reversal invariance of the underlying equations 
of motion \cite{Lindquist01}.

The standard way to solve the problem is to introduce the (backward)
transfer map \cite{Delyon86,Wan90,Li96,Hennig99}
\begin{equation}
u_{n-1} = -v_n + (V_n-\omega + \alpha_n|u_n|^2)u_n, \quad
v_{n-1} = u_n
\label{map}
\end{equation}
where $u_n=\psi_n$ and $v_n=\psi_{n+1}$. Note that these are 
complex quantities therefore the map is nominally four-dimensional.
However, due to conservation of energy and norm, it can be 
reduced to a two-dimensional area-preserving 
map \cite{Delyon86,Wan90,Li96,Hennig99} with 
an additional control parameter (the conserved current $J$). 
The solutions are found by iterating (\ref{map}) 
from the initial point $u_N = T \exp(ikN)$,  $v_N = T \exp(ik(N+1))$
dictated by the boundary conditions of Eq.~(\ref{wave}). 
For fixed $T$ and $k$, the incident and reflected 
amplitudes are determined as 
\begin{equation}
R_0 = \frac{\exp(-ik)u_0-v_0}{\exp(-ik)-\exp(ik)}, \quad
R = \frac{\exp(ik)u_0-v_0}{\exp(ik)-\exp(-ik)}
\nonumber
\end{equation}
and the transmission coefficient is $t(k,|T|^2) = |T|^2/|R_0|^2$
Note that if $(u_0,v_0)=(u_N,v_N)$ (periodic point of the map)
then $t=1$.
 

In order to illustrate the effect in the simplest case we consider 
the DNLS dimer $N=2$. The coefficient $t$ can be thereby computed 
analytically iterating the map twice: 
\begin{equation}
t = \left|\frac{e^{ik}-e^{-ik}}{1+(\nu-e^{ik})(e^{ik}-\delta)}
\right|^2
\end{equation}
where $\delta=V_2 -\omega +\alpha_2T^2$,   
$\nu=V_1-\omega+\alpha_1T^2[1-2\delta\cos k +\delta^2]$.
The formulas applies for $k>0$. As explained above, to solve for 
the case $k<0$ one has to exchange the subscripts $1$ and $2$ which is 
equivalent to reverse the sample. 

\begin{figure}[ht]
\begin{center}
\includegraphics[width=0.4\textwidth,clip]{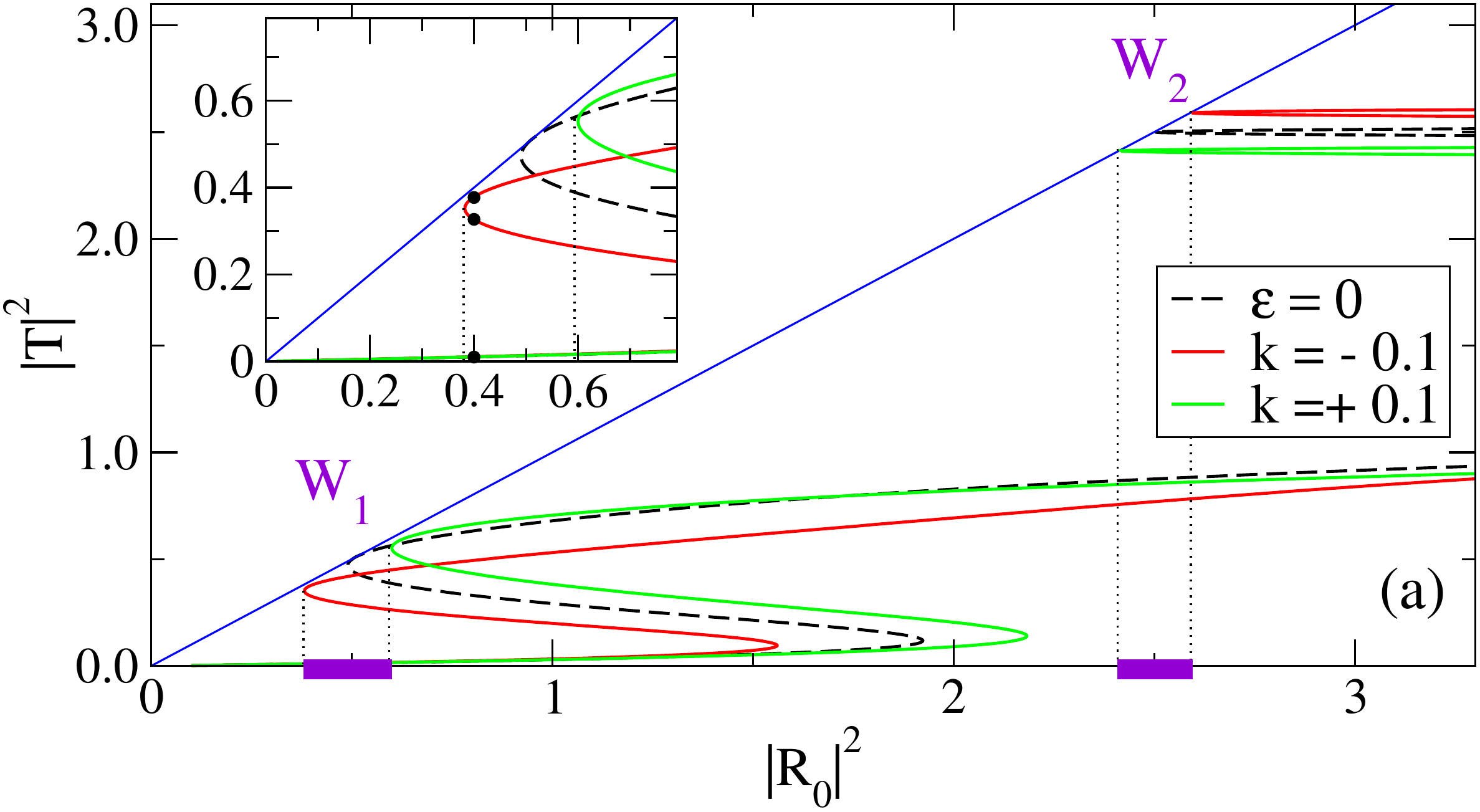}
\includegraphics[width=0.4\textwidth,clip]{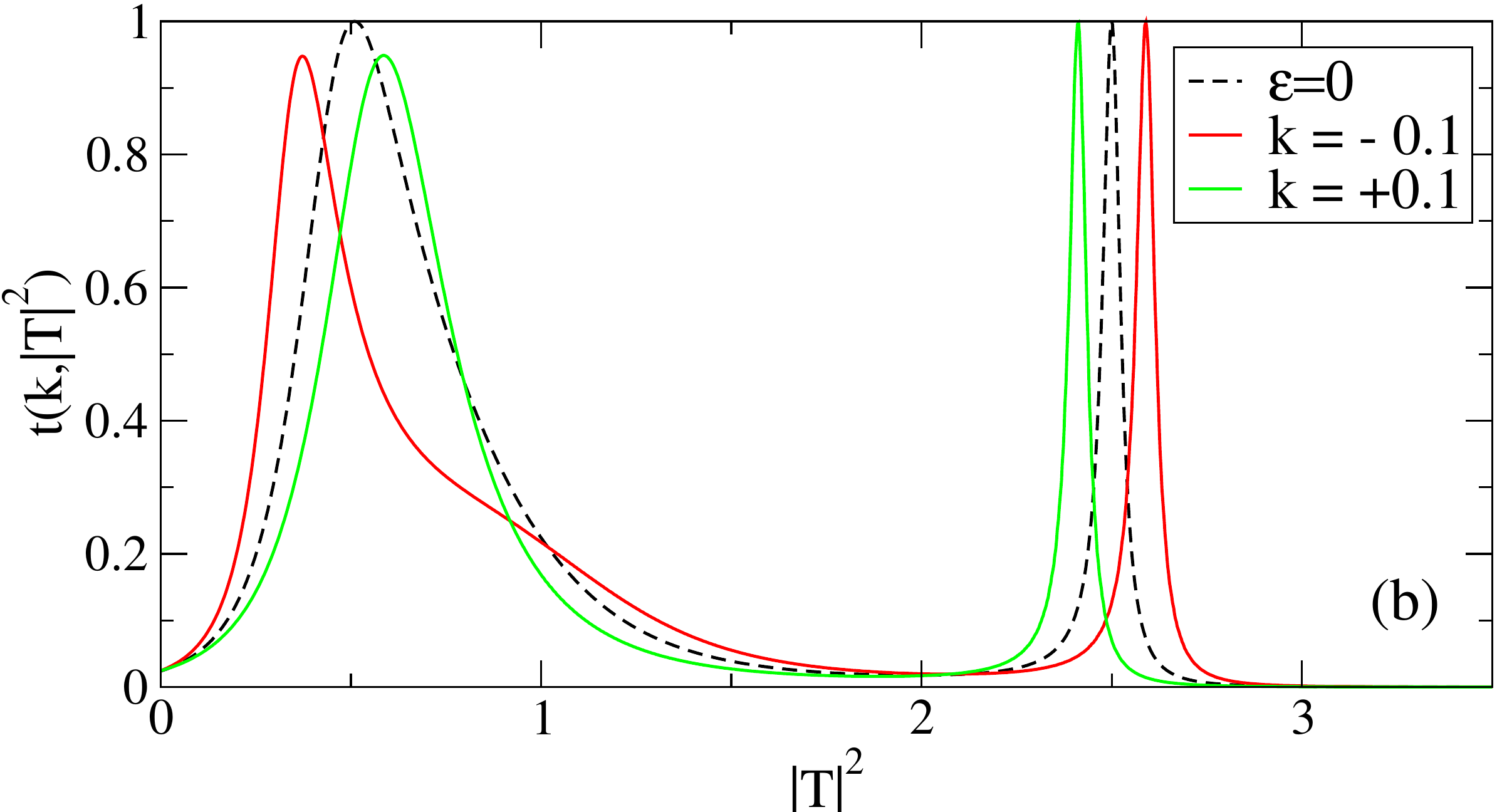}
\caption{(Color online) DNLS dimer, $N=2$, $V_0=-2.5$, $\alpha=1$, $|k|=0.1$.
Comparison between the symmetric case ($\varepsilon=0$, dashed lines)
and the asymmetric one ($\varepsilon=0.05$, solid lines).
(a) Transmission curves, the inset is an enlargement of the 
low-amplitude region. The vertical lines mark the 
turning points of the curves. Accordingly, the heavy lines on the horizontal
axis are the multistability windows $W_{1,2}$ where the diode effect 
maximally occurs. (b) Transmission coefficients as a function of the 
transmitted intensity. 
}
\label{f:fig1}
\end{center}
\end{figure}

Up to now no hypothesis has been made on the coupling coefficients. 
For simplicity, we impose $\alpha_{1,2}=\alpha>0$ henceforth
and let $V_{1,2}=V_0(1\pm\varepsilon)$.
In the symmetric case $\varepsilon=0$, it can be verified that 
transmission is unity for $V_0+\alpha T^2 = 0$ (for $V_0<0$)  
and $V_0+\alpha T^2 =\omega$ (for $V_0<\omega$). These nonlinear resonances can be regarded as the continuation of the
extended  (i.e. perfectly transmitting) states of the corresponding linear
problem \cite{Dunlap90} to nonvanishing $\alpha$.  As a
result, the transmission curves $(|R_0|^2,|T|^2)$ are  tangent to the bisectrix
in two points where $t=1$ (dashed lines in  Fig.~\ref{f:fig1}a and b 
respectively).

\begin{figure}[ht]
\begin{center}
\includegraphics[width=0.4\textwidth,clip]{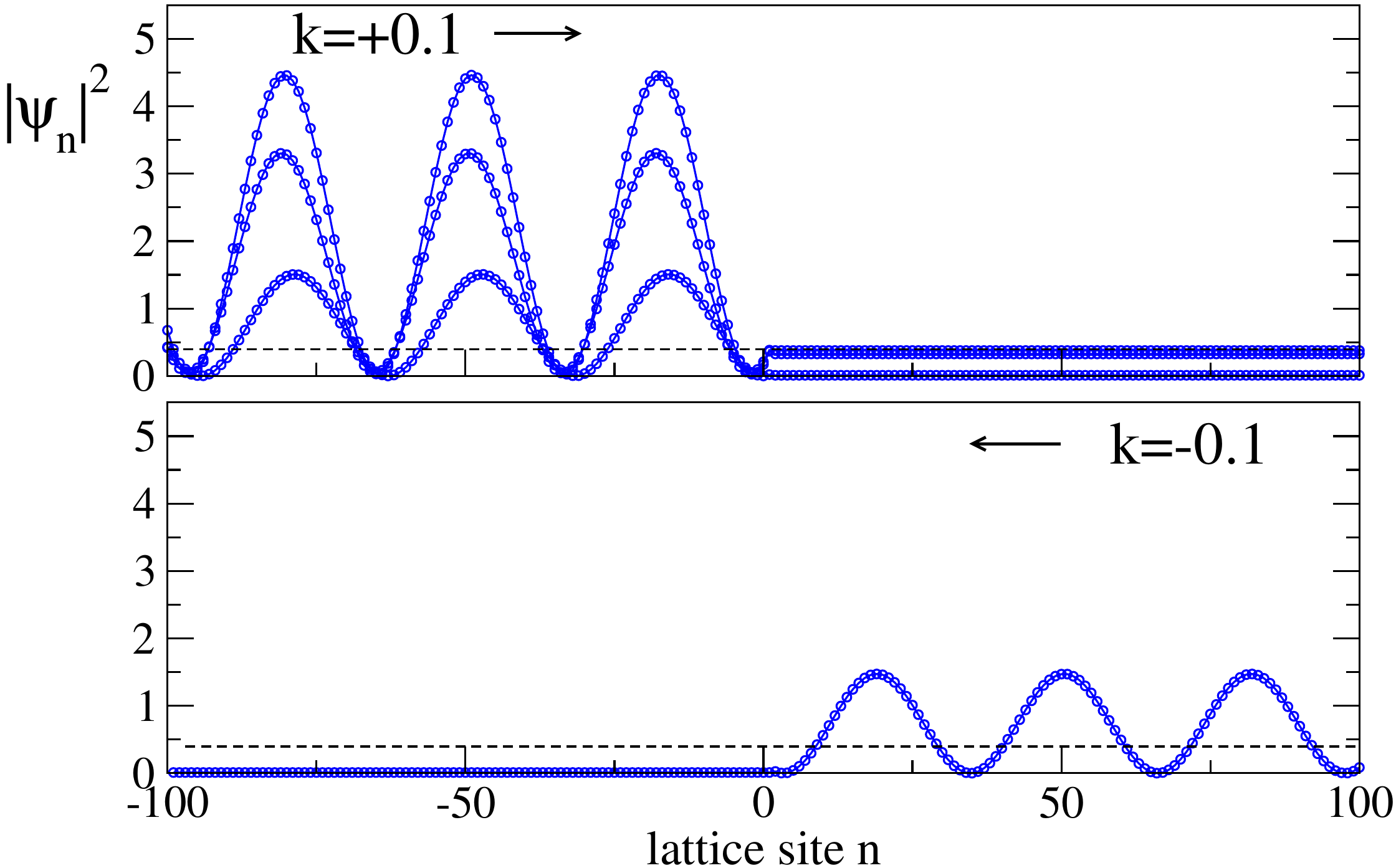}
\caption{(Color online) Square modulus of the solutions corresponding to the 
same incident amplitude $|R_0|^2=0.4$ (marked by full dots 
in the inset of Fig.~\ref{f:fig1}), other parameters 
as in previous figure. 
Upper panel: the three left-propagating 
solutions corresponding to $|T|^2=0.327, 0.377, 0.01$
(top to bottom). Lower panel: the right-propagating solution
corresponding to $|T|^2=0.01$.
The oscillations are caused by the interference between
the incident and reflected waves (wavenumber $2k$).}
\label{f:fig3}
\end{center}
\end{figure}

For the asymmetric case, $\varepsilon \neq 0$, the resonances are 
detuned differently for the $k>0$ and $k<0$ cases, leading to a
nonreciprocal transmission (see the solid lines in Fig.~\ref{f:fig1}a,b)
with maximal $t$ smaller than one.
In particular, there exist two windows $W_{1,2}$ of amplitude of 
the incident wave $R_0$ where we have three solutions for $k>0$ 
and only one for $k<0$. Here the asymmetry become 
maximal. Fig.~\ref{f:fig3} illustrates the different types of solutions 
in the region $W_1$ (see the points in the inset of Fig.\ref{f:fig1}a). 
The two low-amplitude solutions are very similar as expected for 
weak nonlinearity \cite{Miroshnichenko2009}. 

 
\begin{figure}[ht]
\begin{center}
\includegraphics[width=0.55\textwidth,clip]{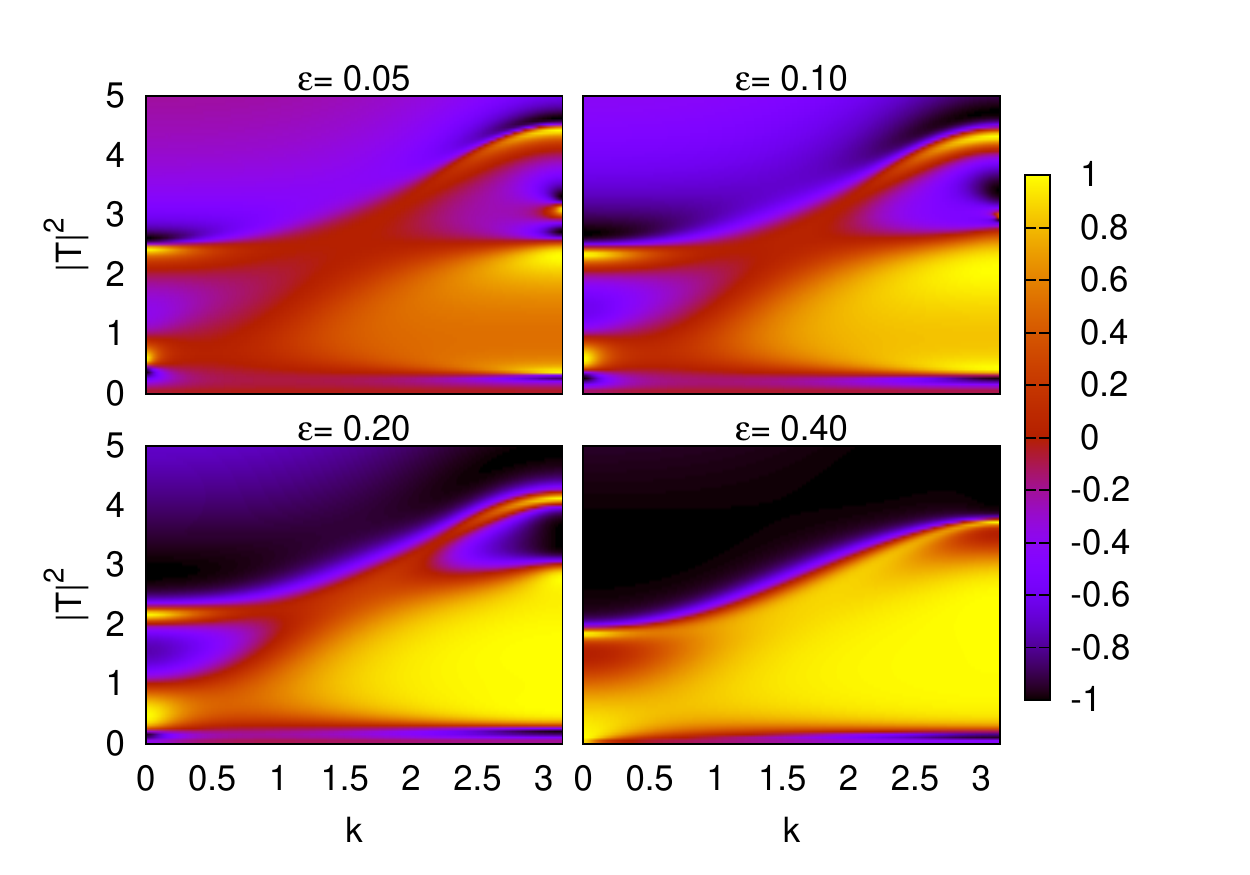}
\caption{(Color online) Contour plot of the rectifying factor (\ref{rfactor})
for increasing asymmetry level $\varepsilon$. Other
parameters as in the previous figure. 
}
\label{f:fig2}
\end{center}
\end{figure}

Since the phenomenon is of nonlinear origin the asymmetry depends
on both frequency and amplitude. To quantify its efficiency, 
in Fig.~\ref{f:fig2} we report the rectifying factor
\begin{equation}
f \;=\; \frac{t(k,|T|^2)-t(-k,|T|^2)}{t(k,|T|^2)+t(-k,|T|^2)}\quad,
\label{rfactor}
\end{equation}
which approaches $\pm 1$ for maximal asymmetry. Note that,
although increasing $\varepsilon$ broadens the regions 
in which $|f|$ is relatively large, the overall transmitted intensity
is reduced as well.   

\begin{figure}[ht]
\begin{center}
\includegraphics[width=0.4\textwidth,clip]{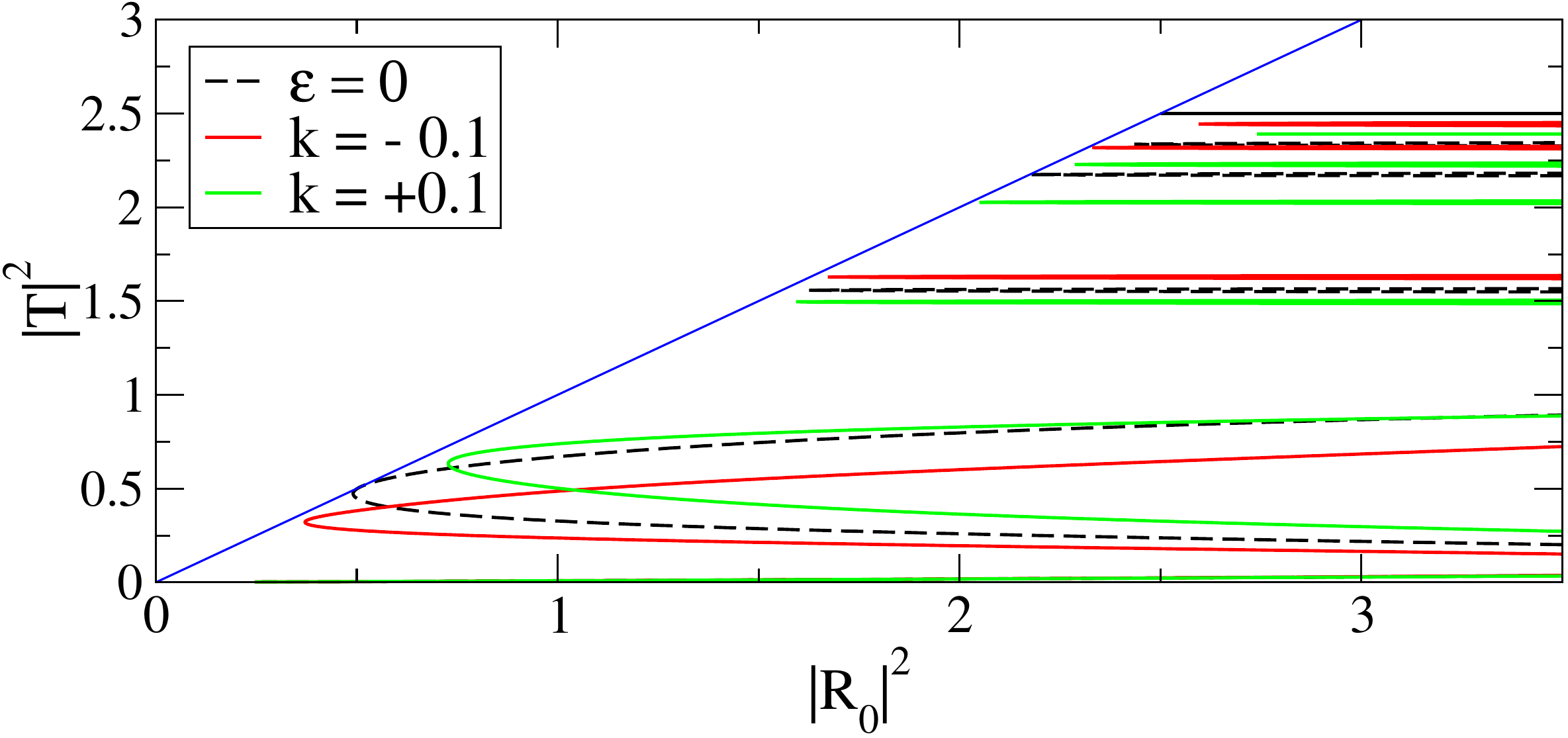}
\includegraphics[width=0.55\textwidth,clip]{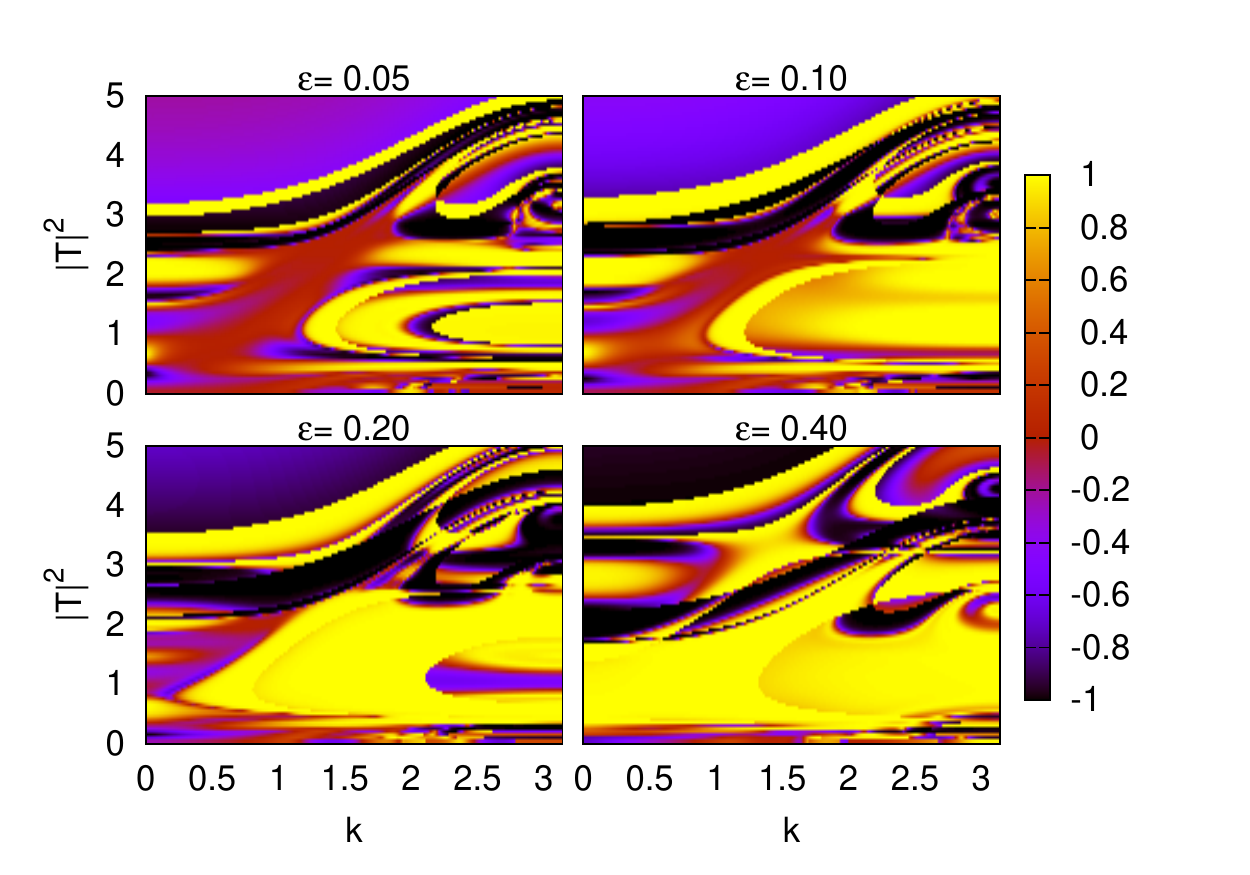}
\caption{(Color online) Upper panel: transmission curves for DNLS with 
$N=4$, linearly-modulated parameter 
$V_n=V_0[1+\varepsilon(1-2(n-1)/(N-1)]$ , other parameters 
as in the previous figures. Upper panel:
comparison between the symmetric case ($\varepsilon=0$, dashed lines)
and the asymmetric one ($\varepsilon=0.05$, solid lines).
Lower panels: contour plots of the rectifying factor (\ref{rfactor})
for increasing asymmetry.
Similar results are obtained for different types of modulation.}
\label{f:fig4}
\end{center}
\end{figure}

Increasing the number $N$ of nonlinear layers will considerably increase the
complexity of the transmission patterns. Indeed,  for large $N$, the
transmitting (resp. nontransmitting)  solutions correspond to bounded (resp.
escaping)  orbits of the map (\ref{map}). Due to the mixed phase-space, $t$ is
nonvanishing only on a fractal set of the parameter  space ~\cite{Delyon86}. The
system thus exhibits a complicated form of multistability, meaning that for a
fixed input one may have (infinitely) many outputs \cite{Delyon86,Hennig99}. We
thus expect that the structure of the windows $W_n$ will approach that of a
Cantor set in the  large $N$ limit. Indeed, Fig.~\ref{f:fig4} indicates that
even for small $N$, the $W_n$ shrink and the texture of the regions  where
$|f|\lesssim 1$ complicates considerably. 

\begin{figure}[ht]
\includegraphics[width=0.5\textwidth,clip]{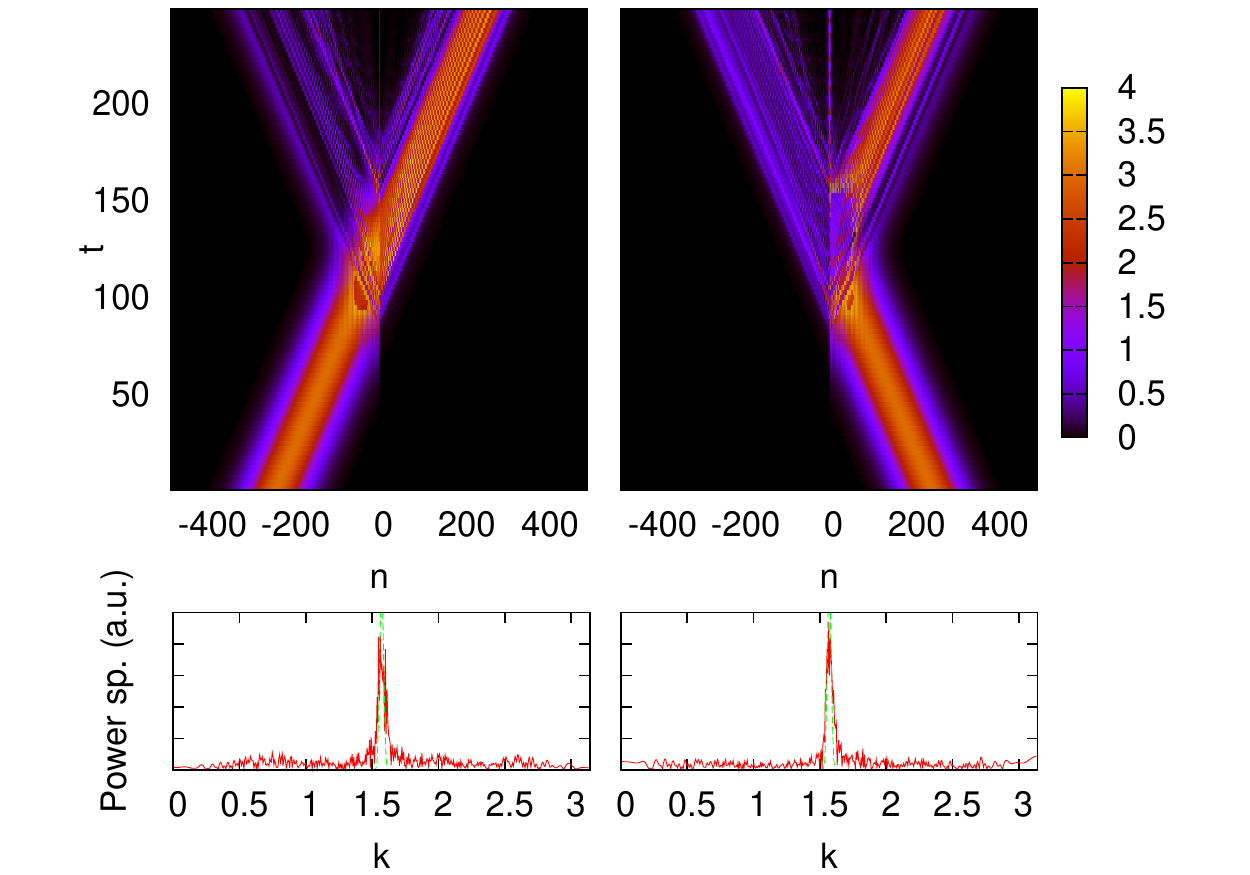}
\caption{(Color online) Numerical simulations of the propagation 
of Gaussian wavepackets, Eq.~(\ref{gauss}) 
impinging on a DNLS dimer. Here $V_0=-2.5$, $|k_0|=1.57$, 
$\varepsilon=0.05$, $M=500$, $|I|^2=3$, $w=10^4$ and 
$n_0=\mp 250$, respectively. 
Lower panels: Power spectra of the real part of $\phi_n$ at 
times $t=0$ (green dashed line) and $t=250$ (red solid line). 
}
\label{f:fig5}
\end{figure}

What are the consequences of the above results on
the transmission of wavepackets? In a nonlinear system
where the superposition principle no longer holds, 
the connection between the two problems is not trivial.
To address this problem, we solved numerically 
the time-dependent DNLS 
\begin{equation}
i \dot \phi_n = V_n\phi_n 
-\phi_{n+1} - \phi_{n-1} + \alpha_n |\phi_n|^2 \phi_n  
\label{tdnls}
\end{equation}
on a finite lattice $|n|\le M$ with open boundary conditions,
for the case of the dimer discussed above. We take
as initial condition a Gaussian wavepacket
\begin{equation}
\phi_n(0)\;=\; I \exp \left[ -\frac{(n-n_0)^2}{w}+ik_0n \right]
\label{gauss}
\end{equation}
The upper panels of Fig.~\ref{f:fig5} display the evolution
of two packets with the same $I$ and opposite wavenumber $k_0$ 
impinging on the nonlinear dimer. The asymmetry of 
their propagation is manifest. Here, the 
parameters have been chosen empirically to obtain 
the maximal asymmetry. More precisely, we first measured the 
wavepacket transmission coefficient $t_{p}$, defined 
as the ratio between the transmitted norm 
$\sum_{n>N}|\phi_n|^2$ at the end of the run divided by the 
initial one $\sum_{n<0}|\phi_n(0)|^2$, as a function of 
$|I|^2$. The data of Fig.~\ref{f:fig5} correspond to 
the amplitude $I$ for which the transmission is maximal
for the left-incoming packet ($t_{p} \simeq 0.8$)
and minimal for the right-incoming one ($t_{p}\simeq 0.3$).
Although the packets are significantly distorted after 
scattering, the Fourier analysis shows that they remain almost 
monochromatic at the incident wavenumber $k_0$
(lower panels of Fig.~\ref{f:fig5}). 
It is also noteworthy that reflection is associated 
with the creation of a localized excitation, strongly
reminiscent of the nonlinear impurity modes 
\cite{Kevrekidis}.

In conclusion, based on the DNLS model, we have demonstrated a mechanism 
which leads to nonreciprocal wave transmission. The new 
class of solutions found here are of interest both theoretically as well as
to envisage possible experimental realizations in nonlinear layered 
photonic or phononic systems. To the extent to which the 
DNLS can be considered a realistic model for such media, 
our results may open the way to novel strategies to control and 
optimize wave propagation and to design devices for sound 
or light rectification.


We thank P.G. Kevrekidis and M. Johansson for fruitful 
discussions. This work is part of the Miur PRIN 2008
project \textit{Efficienza delle macchine termoelettriche: 
un approccio microscopico}. 

\bibliography{diodo,heat}
\bibliographystyle{apsrev}

\end{document}